\documentclass[12pt]{article}

\usepackage{pb-diagram}
\usepackage{latexsym}
\usepackage{amsfonts}

\catcode`\@=11
\def\marginnote#1{}

\newcount\hour
\newcount\minute
\newtoks\amorpm
\hour=\time\divide\hour by60 \minute=\time{\multiply\hour by60
\global\advance\minute by-\hour}\edef\standardtime{{\ifnum\hour<12
\global\amorpm={am}%
        \else\global\amorpm={pm}\advance\hour by-12 \fi
        \ifnum\hour=0 \hour=12 \fi
        \number\hour:\ifnum\minute<10
0\fi\number\minute\the\amorpm}}
\edef\militarytime{\number\hour:\ifnum\minute<10 0\fi\number\minute}

\def\draftlabel#1{{\@bsphack\if@filesw {\let\thepage\relax
   \xdef\@gtempa{\write\@auxout{\string
      \newlabel{#1}{{\@currentlabel}{\thepage}}}}}\@gtempa
   \if@nobreak \ifvmode\nobreak\fi\fi\fi\@esphack}
        \gdef\@eqnlabel{#1}}
\def\@eqnlabel{}
\def\@vacuum{}
\def\draftmarginnote#1{\marginpar{\raggedright\scriptsize\tt#1}}
\def\draft{\oddsidemargin -.5truein
        \def\@oddfoot{\sl preliminary draft \hfil
        \rm\thepage\hfil\sl\today\quad\militarytime}
        \let\@evenfoot\@oddfoot \overfullrule 3pt
        \let\label=\draftlabel
        \let\marginnote=\draftmarginnote

\def\@eqnnum{(\theequation)\rlap{\kern\marginparsep\tt\@eqnlabel}%
\global\let\@eqnlabel\@vacuum}  }

\def\numberbysection{\@addtoreset{equation}{section}
        \def\theequation{\thesection.\arabic{equation}}}

\def\underline#1{\relax\ifmmode\@@underline#1\else
 $\@@underline{\hbox{#1}}$\relax\fi}

\catcode`@=12 \relax

\numberbysection

\topmargin by -\headsep

\def\nonu{\nonumber}
\def\br{\begin{eqnarray}}
\def\er{\end{eqnarray}}
\def\be{\begin{equation}}
\def\ee{\end{equation}}

\def\({\left(}
\def\){\right)}
\def\[{\left[}
\def\]{\right]}

\relax

\def\a{\alpha}

\def\b{\beta}

\def\da{\dagger}

\def\g{\gamma}
\def\G{\Gamma}

\def\l{\lambda}

\def\o{\over}

\def\O{\Omega}
\def\p{\phi}

\def\pa{\partial}

\def\s{\sigma}

\def\t{\tau}
\def\th{\theta}
\def\Th{\Theta}

\def\tp0{\Theta_{+}^{(0)}}
\def\tm0{\Theta_{-}^{(0)}}

\def\cl{{\cal L}}

\def\lie{{\cal G}}

\def\f#1#2#3 {f^{#1#2}_{#3}}

\def\win1{{\sf w_{1+\infty}}}

\def\Win1{{\sf W_{1+\infty}}}

\def\rlx{\relax\leavevmode}
\def\inbar{\vrule height1.5ex width.4pt depth0pt}
\def\IZ{\rlx\hbox{\sf Z\kern-.4em Z}}
\def\IR{\rlx\hbox{\rm I\kern-.18em R}}
\def\IC{\rlx\hbox{\,$\inbar\kern-.3em{\rm C}$}}
\def\IN{\rlx\hbox{\rm I\kern-.18em N}}
\def\IO{\rlx\hbox{\,$\inbar\kern-.3em{\rm O}$}}
\def\IP{\rlx\hbox{\rm I\kern-.18em P}}
\def\IQ{\rlx\hbox{\,$\inbar\kern-.3em{\rm Q}$}}
\def\IF{\rlx\hbox{\rm I\kern-.18em F}}
\def\IG{\rlx\hbox{\,$\inbar\kern-.3em{\rm G}$}}
\def\IH{\rlx\hbox{\rm I\kern-.18em H}}
\def\II{\rlx\hbox{\rm I\kern-.18em I}}
\def\IK{\rlx\hbox{\rm I\kern-.18em K}}
\def\IL{\rlx\hbox{\rm I\kern-.18em L}}
\def\one{\hbox{{1}\kern-.25em\hbox{l}}}
\def\0#1{\relax\ifmmode\mathaccent"7017{#1}%
B        \else\accent23#1\relax\fi}

\def\NPB#1#2#3{{\sl Nucl. Phys.} {\bf B#1} (#2) #3}

\def\PRD#1#2#3{{\sl Phys. Rev.} {\bf D#1} (#2) #3}

\def\JMP#1#2#3{{\sl J. Math. Phys.} {\bf #1} (#2) #3}

\def\LMP#1#2#3{{\sl Letters in Math. Phys.} {\bf #1} (#2) #3}
\def\IJMPA#1#2#3{{\sl Int. J. Mod. Phys.} {\bf A#1} (#2) #3}

\def\TMP#1#2#3{{\sl Theor. Mat. Phys.} {\bf #1} (#2) #3}

\begin{document}

\vspace*{-2 cm}
\noindent

\vskip 1 cm
\begin{center}
{\Large\bf Grassmanian and Bosonic Thirring Models with Jump Defects } \\
\vspace{1 cm}
{A.R. Aguirre, { J.F. Gomes} and  { A.H. Zimerman}}\footnote{  aleroagu@ift.unesp.br, jfg@ift.unesp.br,  zimerman@ift.unesp.br}


{\footnotesize Instituto de F\'\i sica Te\'orica - IFT/UNESP\\
Rua Dr. Bento Teobaldo Ferraz, 271, Bloco II\\
CEP 01140-070, S\~ao Paulo - SP, Brazil}\\

\vspace{1 cm}
{ L.H. Ymai }\footnote{lhymai@yahoo.com.br}


{\footnotesize Universidade Federal do Pampa - UNIPAMPA\\
Rua Carlos Barbosa s/n, Bairro Get\'ulio Vargas,\\
CEP 96412-420, Bag\'e-RS, Brazil.}\\

\vspace{1 cm}

\end{center}

\normalsize
\vskip 0.2cm

\begin{center}
{\large {\bf ABSTRACT}}\\
\end{center}
\noindent
In this paper we discuss the Lax formulation of the Grassmanian and Bosonic Thirring models in the presence of jump defects.
For the Grassmanian case, the defect  is described by B\"acklund transformation which is responsible for preserving the integrability of the model. 
 We then propose an extension of  the B\"acklund transformation for the Bosonic Thirring model which is verified by some B\"acklund transitions like
 Vacuum-One soliton, One soliton - One soliton, One soliton - Two solitons and Two solitons - Two solitons.   The Lax formulation within the space split by the defect leads to the integrability of Bosonic Thirring model.

\section{Introduction}

Recently, there has been some interest in the study of  descontinuities (jump defects) in certain integrable field theories. Such discontinuities connects two solutions  of two different regions.
 B\"acklund transformations  provide the natural setting  for describing integrably such descontinuities since  it relates two  distinct  solutions  of the same equation \cite{corrigan}.  Examples of such situation can be found  for the bosonic \cite{Corr1,Corr2,Bow1,Bow2,Cau,baj,nemes} and supersymmetric \cite{Lean1}, \cite{Lean2} field theories.    In a previous note \cite{Ale} the Grassmanian Thirring model with jump defect was considered by writing  its Lagrangian  density and the conservation of the modified energy and momentum were established.  This last fact has indicated the integrability of the system.  In refs. \cite{Ize}, \cite{Ize2} it was shown using the inverse scattering method, that its solutions were not localized with infinite energy.   Dispite of this fact, we  have reconsidered   its study from the point of view of the Lax pair formulation as an instructive guide to discuss more interesting models such as the Bosonic Thirring model.

In section 2 we introduce the Grassmanian Thirring model, as well as its B\"acklund transformation, and discuss the  defect contribution for the number of particles. This allows us to write the B\"acklund transformation in terms of the modified number of particles.
We then introduce the Lax pair for each side of the defect and show that it is compatible with Coleman's bosonization formula \cite{col}.  We then construct the defect matrix which incorporates the B\"acklund transformation.  We determine a gauge transformation connecting both sides  of the defect.

In section 3, we give the Lax formulation for the Bosonic Thirring model, present the defect contribution for the number of particles and define the B\"acklund transformation by extending the corresponding expression  of the Grassmanian case. The N-soliton solution was presented in \cite{Kuz} by using the inverse scattering method. 
 In subsection 3.1 we present the dressing formulation proposed in \cite{babelon} in 
 order to obtain in a direct way the B\"acklund solutions.  We next show that such formulation naturally determines the defect contribution to the  number of particles and therefore the B\"acklund transformation.
 
 In section 4 we verify several examples  of the obtained solutions describing B\"acklund  transitions: Vacuum-One soliton, One soliton - One soliton, One soliton - Two solitons and Two soliton-Two solitons, obtained by dressing method of section 3.1 .
 
 Finally, in section 5 we write down the modified Lax pair containing the defect and construct a gauge transformation interpolating the two regions.

\section{Grassmanian Thirring Model}

\label{sec:lagran}
The Lagrangian density for Grassmanian Thirring model  can be written as follows,
\begin{eqnarray}
 \cl_p &=& \frac{i}{2}\psi_1^{(p)}(\pa_t - \pa_x)\psi_1^{\dagger(p)} +\frac{i}{2}\psi_1^{\dagger(p)}
 (\pa_t - \pa_x)\psi_1^{(p)} +\frac{i}{2}\psi_2^{(p)}(\pa_t + \pa_x)\psi_2^{\dagger(p)}+\frac{i}{2}\psi_2^{\dagger(p)}
 (\pa_t + \pa_x)\psi_2^{(p)}  \nonumber \\
 &+&  m\big(\psi_1^{(p)}\psi_2^{\dagger(p)} + \psi_2^{(p)}\psi_1^{\dagger(p)} \big) -
 g\big(\psi_1^{\dagger(p)}\psi_2^{\dagger(p)}\psi_2^{(p)}\psi_1^{(p)}\big)\,.
\end{eqnarray}
When considering jump defect at $x=0$ the index $p$ describe the left and right sectors. More explicitly, 
 the lagrangian density for Thirring model describing massive two-component Dirac Grassman fields
$(\psi_1^{(p)},\psi_2^{(p)})$ with $p=1$ corresponding to $x<0$, $p=2$ corresponding to $x >0$, and $g$ is
coupling constant.  The field equations for $x\neq 0$ are given by
\begin{eqnarray}
 i(\pa_t -\pa_x) \psi_1^{(p)} &=& m \psi_2^{(p)} + g\psi_2^{\dagger(p)}\psi_2^{(p)}\psi_1^{(p)}\,,\label{e1}\\
 i(\pa_t + \pa_x) \psi_2^{(p)} &=& m\psi_1^{(p)} + g\psi_1^{\dagger(p)}\psi_1^{(p)}\psi_2^{(p)}\,,\label{e2}\\
 i(\pa_t -\pa_x) \psi_1^{\dagger(p)} &=& -m\psi_2^{\dagger(p)} -g\psi_1^{\dagger(p)}\psi_2^{\dagger(p)}\psi_2^{(p)}\,,\label{e3}\\
 i(\pa_t + \pa_x) \psi_2^{\dagger(p)} &=& -m\psi_1^{\dagger(p)} - g\psi_2^{\dagger(p)}\psi_1^{\dagger(p)}\psi_1^{(p)}\,,\label{e4}
\end{eqnarray}
which are the equations of motion for the Grassmanian Thirring model in the bulk. For $x=0$, the equations corresponding to the defect conditions  are assumed to be \cite{Ale,Ize}, 
\begin{eqnarray}
X&=& (\psi_1^{(2)} + \psi_1^{(1)} ) + {{iag}\over {2m}}\psi_1^{(1)}  X^{\dagger} X \qquad =\,\,\,(\psi_1^{(2)} + \psi_1^{(1)} ) - {{iag}\over {2m}}\psi_1^{(2)}  X^{\dagger} X \nonu \\
&=& ia^{-1}(\psi_2^{(2)}-\psi_2^{(1)}) -\frac{g}{2m}X^\da X \psi_2^{(2)}\,=\,\,\, ia^{-1}(\psi_2^{(2)}-\psi_2^{(1)}) -\frac{g}{2m}X^\da X \psi_2^{(1)},\label{eq4.8}
\end{eqnarray}
its respective hermitian conjugated equations
\begin{eqnarray}
X^\da &=& (\psi_1^{\da(2)} + \psi_1^{\da(1)} ) - {{iag}\over {2m}}\psi_1^{\da(1)}  X^{\dagger} X \qquad \,\,\,\,\, =\,\,\,(\psi_1^{\da(2)} + \psi_1^{\da(1)} ) + {{iag}\over {2m}}\psi_1^{\da(2)}  X^{\dagger} X  \\
&=& -ia^{-1}(\psi_2^{\da(2)}-\psi_2^{\da(1)}) -\frac{g}{2m}X^\da X \psi_2^{\da(2)}\,=\,\,\, -ia^{-1}(\psi_2^{\da(2)}-\psi_2^{\da(1)}) -\frac{g}{2m}X^\da X \psi_2^{\da(1)},\qquad \mbox{}
\end{eqnarray}
and the time derivatives
\begin{eqnarray}
\pa_t X &=& {{m}\over {2a}} (\psi_1^{(2)} - \psi_1^{(1)}) - {{im}\over {2}} (\psi_2^{(2)} + \psi_2^{(1)}) \nonumber \\
             &-& \frac{ig}{4}\left[\psi_1^{\da(1)}\psi_1^{(1)} + \psi_1^{\da(2)}\psi_1^{(2)}  + \psi_2^{\da(1)}\psi_2^{(1)} + \psi_2^{\da(2)}\psi_2^{(2)}\right] X \\
\pa_t X^{\dagger}  &=& {{m}\over {2a}} (\psi_1^{\dagger(2) } - \psi_1^{ \dagger(1) }) + {{im}\over {2}} (\psi_2^{ \dagger(2) } + \psi_2^{ \dagger(1) })\nonumber \\
&+& \frac{ig}{4}\left[\psi_1^{\da(1)}\psi_1^{(1)} + \psi_1^{\da(2)}\psi_1^{(2)}  + \psi_2^{\da(1)}\psi_2^{(1)} + \psi_2^{\da(2)}\psi_2^{(2)}\right] X^\da \label{eqn4.13}
\end{eqnarray}
These equations correspond precisely to  the B\"acklund transformations for the classical Grassmanian Thirring model \cite{Ize}. 
The first integral of motion to be considered is the \emph{number of particles} $N$ given by ,
\br
 N &=& \int_{-\infty}^0 dx \left(\psi_1^{\dagger(1)}\psi_1^{(1)} + \psi_2^{\da(1)}\psi_2^{(1)} \right) + \int_{0}^{\infty} dx \left(\psi_1^{\da(2)}\psi_1^{(2)} + \psi_2^{\da(2)}\psi_2^{(2)} \right).
\er
Using the equations of motion (\ref{e1})-(\ref{e4}), we have
\br
 \frac{dN}{dt} &=& \left[\psi_1^{\da(1)}\psi_1^{(1)} - \psi_2^{\da(1)}\psi_2^{(1)} \right]\bigg|_{x=0} - \left[\psi_1^{\da(2)}\psi_1^{(2)} - \psi_2^{\da(2)}\psi_2^{(2)} \right]\bigg|_{x=0}.
\er
From   the B\"acklund transformation and equations of motion we find the modified conserved number of particles, 
\br
{\cal N} &=& N + \frac{a}{m}X^\da X = N + N_D, \qquad \qquad
{\rm where} \qquad \qquad 
N_D  = \frac{a}{m}X^\da X\label{eqn2.13}
\er
 corresponds to the defect contribution to the number of particles.
Since $N_D^2 = 0$, (\ref{eq4.8}) can be written as
\br
X &=& \psi_1^{(1)} e^{{{{ig}\o {4}}}N_D} + \psi_1^{(2)} e^{-{{{ig}\o {4}}}N_D} 
= ia^{-1} \( \psi_2^{(2)} e^{{{{ig}\o {4}}}N_D} -\psi_2^{(1)} e^{-{{{ig}\o {4}}}N_D} \). 
\label{nd}
\er
The defect contribution to the number of particle quantity $N_D$ is the simplest modified conserved charge and turns out to be very important for describing the B\"acklund transformation in an alternative form. The arguments for the integrability of the Grassmaniann Thirring model have already been examined in a previous work \cite{Ale}, by calculating the respective modified conserved energy and momentum quantities.


\subsection{Lax Pair}

So far we have studied the classical integrability of the Grassmanian Thirring model
with jump defect by constructing the lowest conserved quantity, namely, the modified number of particles.  The integrability of the model involves also higher conservation laws  which are encoded within the Lax pair formalism within the $\hat {sl}(2,1)$ affine Lie algebra and principal gradation (see appendix C), as follows
\br
 {\cal L}_t &=& i\pa_t + A_t\,,\qquad {\cal L}_x\,=\,\,=\, i\pa_x + A_x\,,
\er
where the connections $A_t$ and $A_x$ take the following forms
\begin{eqnarray}
 A_x &=& \frac{g}{2}\big(\psi_2^{\dagger}\psi_2-\psi_1^{\dagger}\psi_1\big)h_1 + \frac{m}{2}\left(\l-{\l}^{-1} \right)\big(h_1+2h_2\big)
-\a\big(\psi_1^\da\l^{1/2}+\psi_2^\da\l^{-1/2}\big) E_{-(\a_1+\a_2)}\nonu \\ 
&+& \a\big(\psi_1\l^{1/2}-\psi_2\l^{-1/2} \big)E_{-\a_2} -\a\big(\psi_1\l^{1/2} + \psi_2 \l^{-1/2}\big) E_{\a_1+\a_2} +\a\big(\psi_1^\da\l^{1/2} -\psi_2^\da \l^{-1/2}\big) E_{\a_2} \nonu \\
&\mbox{}&\label{eq4.33} \\
A_t &=& -\frac{g}{2}\big(\psi_1^{\dagger}\psi_1+\psi_2^{\dagger}\psi_2\big)h_1 + \frac{m}{2}\left({\l}^{-1}+\l \right)\big(h_1+2h_2\big)
        -\a\big(\psi_1^\da\l^{1/2} -\psi_2^\da\l^{-1/2}\big)E_{-(\a_1+\a_2)} \nonu \\
     &+&\!\! \a\big(\psi_1\l^{1/2}+\psi_2\l^{-1/2} \big)E_{-\a_2} -\a\big(\psi_1 \l^{1/2}-\psi_2\l^{-1/2}\big)E_{\a_1+\a_2}+\a\big(\psi_1^\da\l^{1/2}+\psi_2^\da\l^{-1/2}\big) E_{\a_2},\nonu \\
\label{eq4.34} 
\end{eqnarray}
\noindent where we have introduced the parameter $\a=\sqrt{\frac{mg}{2}}$. From the zero curvature condition $\left[{\cal L}_t,{\cal L}_x\right]=0$, we immediately find the equations of motions (\ref{e1})-(\ref{e4}), and the equation corresponding to zero grade in $\l$ leads to the following compatibility equation,
\br
 i(\pa_t -\pa_x)(\psi^\dagger_1\psi_1) - i(\pa_t + \pa_x)(\psi_2^\dagger\psi_2) +2m\big(\psi_2\psi_1^\dagger + \psi_2^\dagger\psi_1\big) &=& 0 \,,
\er
which can be rewiten as
\br
 i\pa_t(\psi^\dagger_1\psi_1-\psi_2^\dagger\psi_2) - i\pa_x(\psi^\dagger_1\psi_1+\psi_2^\dagger\psi_2) +2m\big(\psi_2\psi_1^\dagger + \psi_2^\dagger\psi_1\big) &=& 0 .
\er
Then, remembering that the fermionic currents $j^\mu = \bar{\psi}\gamma^\mu\psi$ in component fields are given by
\br
j^0 &=& \psi^\dagger_1\psi_1+\psi_2^\dagger\psi_2, \qquad {\rm and} \qquad j^1 \,=\, -\psi^\dagger_1\psi_1+\psi_2^\dagger\psi_2,
\er
we have
\br
 -i\pa_t j^1 - i\pa_x j^0 + 2m\big(\psi_2^\dagger\psi_1-\psi_1^\dagger\psi_2  \big) &=&0\,.
\er
Now, by applying the Coleman's bosonization rules \cite{col}
\br
 j^0 &=& \frac{\beta}{2\pi}\pa_1\varphi\,\qquad j^1 \,=\,-\frac{\b}{2\pi}\pa_0\varphi\,,
\er
and using the Mandelstam's operators \cite{Mandel}
\br
 \psi_2^\dagger\psi_1 &=& \frac{\hbar}{2\pi\Lambda}: e^{i\b\varphi}: \qquad \psi_1^\dagger\psi_2\,=\, \frac{\hbar}{2\pi\Lambda}: e^{-i\b\varphi}:
\er
we obtain
\br
 \big(\pa_t^2 -\pa_x^2\big)\varphi &=& -\frac{4m_ 0}{\b}\sin\b\varphi\,, 
\er
the sine-Gordon equation, where the masses of the two theories are related by
\br
 m\left(\frac{\hbar}{\Lambda}\right) &=& m_0\,,
\er
where $\Lambda$ is a cut-off introduced\ by Mandelstam.


\subsection{Defect Matrix for the Grassmannian Thirring model}
The Thirring Model can be described by the Lax pair (\ref{eq4.33}) and (\ref{eq4.34}). In the presence of a defect, the integrability of the system is studied by splitting the space into two overlapping regions, $x\leq b$ and $x\geq a$ with $a<b$. Inside the overlap region, $a\leq x \leq b$, we define the Lax pair to be
\begin{eqnarray}
 \hat{A}_t^{(1)} &=& A_t^{(1)} -\theta(x-a)\left[\left\{X-(\psi_1^{(2)} + \psi_1^{(1)}) - \frac{iag}{2m}X^\da X \psi_1^{(1)} \right\} E_{(\a_1 +\a_2)}\right.\nonumber \\
 &\mbox{}&\left. +\left\{X^\dagger - (\psi_1^{\dagger(2) } + \psi_1^{\dagger(1)}) + \frac{iag}{2m} X^\da X \psi_1^{\dagger(1)} \right\} E_{-(\a_1 +\a_2)}\right] \\[0.2cm]
 \hat{A}_x^{(1)} &=& \theta(a-x)A_x^{(1)},\\[0.2cm]
 \hat{A}_t^{(2)} &=& A_t^{(2)} - \theta(b-x)\left[\left\{X -ia^{-1}(\psi_2^{(2)} - \psi_2^{(1)}) +\frac{g}{2m}X^\da X\psi_2^{(1)} \right\}E_{\a_2} \right. \nonumber \\
 &\mbox{}& \left. +\left\{X^\dagger + ia^{-1}(\psi_2^{\dagger(2)} - \psi_2^{\dagger(1)}) +\frac{g}{2m}X^\da X\psi_2^{\dagger(1)} \right\} E_{-\a_2} \right] ,\\[0.2cm]
 \hat{A}_x^{(2)} &=& \theta(x-b) A_x^{(2)} .
\end{eqnarray}
\noindent Within the overlap region, the Lax pair denoted by suffices $p=1,2$, are related by a gauge transformation. Thus, we have
\begin{eqnarray}
 i\pa_t K &=& K\hat A_t^{(2)}(t,b)-\hat A_t^{(1)}(t,a)K\,. \label{eq2.40}
\end{eqnarray}
We will assume that the matrix $K$ can be written by the following $\l$-expansion,
\br
 K &=& K_{-1/2} +  K_0 + K_{1/2} ,
\er
where $K_i$ corresponds to an element of grade $\l^{i}$. As usual, the entries $K_{ij}$ are calculated grade by grade in a $\l$-expansion of the equation (\ref{eq2.40}). After some matricial computations,  a consistent solution for the defect matrix $K$ is then given in the following form, \\
{\footnotesize{
\br
 K \!\!\!\!&=& \!\!\!\!\!\left[\!\!
   \begin{array}{ccc} \l^{1/2}e^{-\frac{iga}{2m}X^\da X} -\l^{-1/2}(ia^{-1})e^{\frac{iga}{2m}X^\da X} & 0 & \sqrt{\frac{2g}{m}}X\\[0.3cm]
   0 & \l^{1/2}e^{\frac{iga}{2m}X^\da X} -\l^{-1/2}(ia^{-1})e^{\frac{-iga}{2m}X^\da X}  & -\sqrt{\frac{2g}{m}}X^\da\\[0.3cm] 
   \sqrt{\frac{2g}{m}}X^\da & -\sqrt{\frac{2g}{m}}X & -\l^{1/2} - \l^{-1/2}(ia^{-1}) 
   \end{array}\!\!\!
 \right]\nonumber\\
\er}}
\noindent where the exponentials can be expanded as
\br
 e^{\pm \frac{iga}{2m}X^\da X} &=& 1 \pm \frac{iga}{2m}X^\da X,
\er
and the auxiliary fields $X$ and $X^\da$ satisfy the equations (\ref{eq4.8})-(\ref{eqn4.13}) corresponding to the B\"acklund transformations for the Grassmanian Thirring model.


\section{Lax Formulation of the Bosonic Thirring Model}

Now for considering the Bosonic Thirring model, we start from the zero curvature representation using the
following Lax pair in the light-cone coordinates\footnote{Here, we use the light-cone coordinates $\xi=\frac{1}{2}(t+x)$, $\eta=\frac{1}{2}(t-x)$.},
\br
{\cal L}_{\xi} &=& i\pa_\xi + A_{\xi}, \qquad 
 {\cal L}_{\eta} = i\pa_\eta +A_{\eta} ,
\er 
with
\br
 A_{\xi} &=& \sqrt{mg}\,\phi_1\, {\sigma}_+^ {(0)} - \sqrt{mg}\,\phi_1^{\dagger}\, {\sigma}_-^ {(+1)} -\frac{g}{2}(\phi_1^\dagger \phi_1)\sigma_3^{(0)} -\frac{m}{2}\,\s_3^{(+1)} + \a_+\,\mathcal{C}, \nonumber \\
 A_{\eta} &=& - \sqrt{mg}\,\phi_2\, {\sigma}_+^ {(-1)} + \sqrt{mg}\,\phi_2^{\dagger}\, {\sigma}_-^ {(0)} -\frac{g}{2}(\phi_2^\dagger \phi_2)\sigma_3^{(0)} -\frac{m}{2}\,\s_3^{(-1)} + \a_-\,\mathcal{C},
 \label{lax}
\er
where $\phi_i$ are commuting fields and the Lax connections take values in the $\hat{s}l(2)$-affine
Kac-Moody algebra $\widehat{\mathcal{G}}$ defined by the commutations
relations 
\br
 \left[ \s_3^{(m)},\s_3^{(n)}\right] &=& 2m\mathcal{C}\delta _{m+n,0},  \nonumber \\
\left[ \s_3^{(m)},\s_{\pm }^{(n)}\right] &=&\pm 2\s_{\pm }^{(m+n)},  \nonumber \\
\left[ \s_{+}^{(m)},\s_{-}^{(n)}\right] &=& \s_3^{(m+n)}+m\mathcal{C}\delta _{m+n,0},  \nonumber \\
\left[ d,T^{(n)}\right] &=&nT^{(n)},\qquad T^{(n)}\equiv \s_3^{(m)},
\s_{\pm }^{(n)}. 
\er 
where $\mathcal{C}$ represents the central term.
The principal grading for the  $\hat{s}l(2)$ is generated by the operator 
\br
 Q &=& 2d + \frac{1}{2} \s_3^{(0)}.
\er 
The  grading operator $Q$ decomposes the algebra
$\widehat{\mathcal{G}}$ into subalgebras generated by elements of
positive, negative and zero grades respectively, 
\br
\widehat{\mathcal{G}} &=& \widehat{{\cal G}}_{+}\oplus
\widehat{{\cal G}}_{0}\oplus \widehat{{\cal G}}_{-}. 
\er 
From the zero curvature condition $\big[{\cal L}_{\xi},{\cal L}_{\eta}\big] =
0 $, we get the field equations for the Bosonic Thirring model 
\br
i\pa_\eta\phi_1 &=& m \phi_2 + g(\phi_2^{\dagger}\phi_2)\phi_1\,,\label{eq2.29}\\
 i\pa_\xi \phi_2 &=& m\phi_1 + g(\phi_1^{\dagger}\phi_1)\phi_2\,,\\
 i\pa_\eta \phi_1^{\dagger} &=& -m\phi_2^{\dagger} -g(\phi_2^{\dagger}\phi_2)\phi_1^{\dagger}\,,\\
 i\pa_\xi  \phi_2^{\dagger} &=& -m\phi_1^{\dagger} - g(\phi_1^{\dagger}\phi_1)\phi_2^{\dagger} \,,\label{eq2.32}
\er 
together with the following equations
\br
 i\pa_\xi(\phi_2^{\dagger}\phi_2) -  i\pa_\eta(\phi_1^{\dagger}\phi_1) - 2m(\phi_1^{\dagger}\phi_2 -\phi_2^{\dagger}\phi_1) &=& 0,\label{1.10}\\
 i\pa_\xi \a_-  - i\pa_\eta \a_+ +mg(\phi_1^{\dagger}\phi_2) + \frac{m^2}{2} & =& 0 .\label{1.11}
\er 
 Equation (\ref{1.10}) is a straightforward consequence of the field equations (\ref{eq2.29})-(\ref{eq2.32}), and the Eq.(\ref{1.11}) determines the dependence of the fields $\a_{\pm}$ in terms of the massive fields
$\phi$'s.
The conserved quantity which gives the number of particles in the bulk is given by
\br
 N = \int_{-\infty}^{\infty} dx\,\[\p_1^\da\p_1 + \p_2^{\da}\p_2 \]\,,
\er
Let $N_D$ be  the defect contribution to the number of particles density, which satisfies
\br
 \frac{dN_D}{dt} &=& -\[\p_1^{\da(1)}\p_1^{(1)} - \p_2^{\da(1)}\p_2^{(1)} \]\bigg|_{x=0} + \[\p_1^{\da(2)}\p_1^{(2)} - \p_2^{\da(2)}\p_2^{(2)} \]\bigg|_{x=0}.
 \label{3.13}
\er
 The auxiliary functions $X$ defining the B\"acklund transformation will be assumed, according to (\ref{nd}), to  satisfy the following algebraic relations,
\br
X \!\!\!\!&=& \!\!\!\! \p_1^{(1)} \exp\[\frac{ig N_D}{4}\] + \p_1^{(2)} \exp\[\frac{-ig N_D}{4}\]  \,=\,\frac{i}{a}\[\p_2^{(2)} \exp\[\frac{ig N_D}{4}\] - \p_2^{(1)} \exp\[\frac{-ig N_D}{4}\] \], \nonumber\\
&\mbox{}&\label{eq3.14}
\er
and
\br
 X^{\dagger} \!\!\!\!&=&\! \!\!\! \p_1^{\dagger (1)} \exp\[-\frac{ig N_D}{4}\] + \p_1^{\dagger (2)} \exp\[\frac{ig N_D}{4}\]  \,=\, \frac{-i}{a}\left[\p_2^{\dagger (2)} \exp\[\frac{-ig N_D}{4}\right] - \p_2^{\dagger (1)}  \exp\[\frac{ig N_D}{4}\] \] \nonumber\\ 
\label{s3e47}
\er
 As a consequence of (\ref{eq2.29})-(\ref{eq2.32}) in (\ref{s3e47}), we also can obtain the following relations,
\br
 \pa_{\xi} X &=&\!\! ma^{-1}\(\p_1^{(2)} \exp\[\frac{ig N_D}{4}\] - \p_1^{(1)} \exp\[\frac{-ig N_D}{4}\]   \) -\frac{ig}{2}\(\p_1^{\da(1)}\p_1^{(1)} +\p_1^{\da(2)}\p_1^{(2)} \)X \quad \mbox{}\label{s3e53}\\[0.3cm]
 \pa_{\eta} X &=&\!\! -im\(\p_2^{(1)} \exp\[\frac{ig N_D}{4}\] + \p_2^{(2)} \exp\[\frac{-ig N_D}{4}\]  \) -\frac{ig}{2}\(\p_2^{\da(1)}\p_2^{(1)} +\p_2^{\da(2)}\p_2^{(2)} \)X. \qquad\quad\mbox{} \label{s3e54}
\er
The integrability condition for the above equations, 
\br
\pa_\xi(\pa_\eta X) =\pa_\eta(\pa_\xi X),
\er
implies that both $\p_i^{(1)}$ as well as $\p_i^{(2)}$ are solutions of the equations of motion (\ref{eq2.29})-(\ref{eq2.32}) for the Bosonic Thirring model. Then, we identify the eqs. (\ref{s3e47})-(\ref{s3e54}) to be the \emph{auto-B\"acklund} \emph{transformations} for the Bosonic Thirring model. These transformations have shown to be totally consistent.
Using Eqs. (\ref{eq3.14})-(\ref{s3e54}) we found that the defect contribution for the number of particles in this case can be written as,
\br
 N_D &=&\frac{2}{g}\arcsin\[\frac{ga}{2m}X^\da X\] .\label{eq3.19}
\er
Notice that taking a naive Grassmaniann limit, $N_D$ given by Eq. (\ref{eq3.19}) is consistent with the expression given in Eq. (\ref{eqn2.13}).

\subsection{Dresssing Formulation}

We now present a systematic way of constructing
solutions by the dressing procedure of refs. \cite{babelon}.  Its 
advantage is to  provide a direct connection between certain quantities naturally appearing within the dressing  formalism and  in
 the B\"acklund transformation as we shall see in (\ref{ref1}) and (\ref{ref2}).  The key ingredient  is the existence of two
gauge transformations $\Theta_+ = \exp ({\lie_{\geq}})  $ and $\Theta_- = \exp({\lie_\leq}) $
mapping the vacuum in a non-trivial configuration, i.e., 
\br
 A_{\mu}^{vac} \longrightarrow A_\mu \equiv \Theta_{\pm}^{-1}i\pa_\mu \Theta_{\pm} + \Theta_{\pm}^{-1}A_\mu^{vac} \Theta_{\pm}, \qquad \mu = \{\eta,\xi \},\label{eq2.35}
\er 
As consequence of the graded structure, the form of the Lax connection is preserved
by these transformations. Since $A_\mu$ and $A_{\mu}^{vac}$ 
satisfy the zero curvature condition, they are of the  form
\br
 A_\mu = iT\pa_\mu T^{-1},\qquad A_\mu^{vac} = iT_0\pa_\mu T_0^{-1}
\er 
where $T$ and $T_0$ are  group elements. From the equivalence of the two dressing transformations (\ref{eq2.35}) we find that 
\br
 \Theta_-\Theta_+^{-1} &=& T_0 \rho T_0^{-1},\label{2.3}
\er
where $\rho$ is a constant group element. In order to construct sistematically  soliton solutions we now define the vacuum configuration, 
\br
 \phi_1^{(0)} =\phi_2^{(0)} = \phi_1^{\dagger (0)} = \phi_2^{\dagger (0)} = 0, \qquad \a_+^{(0)}=\frac{-im^2\eta}{4},\qquad  \a_-^{(0)} = \frac{im^2\xi}{4}.\label{eq2.38}
\er 
and Lax connections (\ref{lax}) become
\br
 A_\xi^{vac} =  -\frac{m}{2}\,\s_3^{(+1)} + \frac{-im^2\eta}{4}\,\mathcal{C}, \qquad A_\eta^{vac} = -\frac{m}{2}\,\s_3^{(-1)} + \frac{im^2\xi}{4}\mathcal{C}.
\er 
They are associated to the following linear problem 
\br
 i\pa_\eta T_0 &=& -A_\eta^{vac} T_0,\qquad
  i\pa_\xi T_0 = -A_\xi^{vac} T_0,
\er 
which is solved as follows
\br
 T_0 &=& e^{-i\eta E^- -i\xi E^+}, \qquad {\rm with} \qquad E^{\pm} \equiv \frac{m}{2}\,\s_3^{(\pm 1)}.
\er 
The dressing matrices $\Th_{\pm}$ are now determined  by the gauge transformation (\ref{eq2.35}) with
\br
 \Th_+ &=& e^{m(0)}e^{m(1)}e^{m(2)}\cdots \quad\quad
 \Th_- = e^{l(0)}e^{l(-1)}e^{l(-2)}\cdots
\er 
where $\Th_+$ is constructed from elements $m(k)$ of a subalgebra containing grade $k \geq 0$, while $\Th_-$ is constructed from elements $l(k)$ of a subalgebra containing grade $k\leq 0$. From eq.(\ref{eq2.35}) we get the following results for the first few elements $m(k)$ and $l(k)$ 
\br
 m(0) &=& \chi_+ \s_3^{(0)} + \nu_+ \mathcal{C}, \qquad l(0) = (i\pi-\chi_+)\s_3^{(0)} + \nu_- \mathcal{C},\\
 m(1) &=& \sqrt{\frac{g}{m} }\left[\phi_2 \s_+^{(0)} + \phi_2^{\dagger}\s_-^{(1)}\right],\qquad l(-1) = -\sqrt{\frac{g}{m} }\left[\phi_1^\da \s_-^{(0)} + \phi_1\s_+^{(-1)}\right]\\
 m(2) &=& a_+ \s_3^{(1)}, \qquad l(-2) = a_-\s_3^{(-1)},
\er
where the fields $\phi$'s satisfy the equations of motion(\ref{eq2.29})-(\ref{eq2.32}), and
the fields $\chi_+, \nu_{\pm}, a_{\pm}$ satisfy the following
equations, 
\br
\hspace{-1cm} i\pa_\xi \chi_+ &=& -\frac{g}{2}(\phi_1^{\dagger}\phi_1), \!\qquad \qquad i\pa_\eta \chi_+ = \frac{g}{2}(\phi_2^{\dagger}\phi_2),\label{eq2.47}\\
 i\pa_\xi \nu_+ &=& \a_+ -\a_+^{(0)},\qquad \qquad  \,  i\pa_\eta \nu_+ = \a_- -\a_-^{(0)} - ma_+ - \frac{g}{2}(\phi_2^{\dagger}\phi_2),\\
 i\pa_\eta \nu_- &=& \a_- -\a_-^{(0)}, \qquad \qquad i\pa_\xi \nu_- \,=\,\a_+ -\a_+^{(0)} +ma_- -\frac{g}{2}(\phi_1^{\dagger}\phi_1).
\er 
The $(\xi,\eta)$-dependence of the fields is given explicitly
by the right-hand-side of Eq.(\ref{2.3}). In fact, the solutions can
be calculated by taking the expectation value between states of a
given representation of $\widehat{\cal{G}}$. As usual, we consider
the highest weight representation of the $\hat{s}l(2)$. Firstly, let
$|\l_0\rangle$ and $|\l_1\rangle$ be the corresponding highest
weight states of $\hat{s}l(2)$, and define the $\tau$-functions as
follows 
\br
 \tau_i &=& \langle \l_i | \Theta_-\Theta_+^{-1} |\l_i \rangle  = \langle \l_i |T_0 \rho T_0^{-1} |\l_i \rangle , \qquad i=1,2.\label{eq2.50}
\er 
The soliton solutions are obtained by choosing the constant element $\rho = e^{V}$, as the exponential of an eigenvector $V$ of the elements of algebra $E^{\pm}$. This eigenvector can be constructed in the following way
\br
 V_{\pm}(\g) = \sum\limits_{n \in \mathbb{Z}} \g^{-n} \s_{\pm}^{(n)},
\er 
satisfying the following commutation relations \br
 \big[E^{+}, V_{\pm}(\g) \big] &=& \pm m \g\, V_{\pm}(\g),\label{eq2.52}\\
 \big[E^{-}, V_{\pm}(\g) \big] &=& \pm \frac{m}{\g}\, V_{\pm}(\g).\label{eq2.53}
\er 
Its clear from (\ref{eq2.52})-(\ref{eq2.53}) that $V_+(\g)$ and $V_-(-\g)$ have the same eigenvalue. From (\ref{2.3}) we obtain 
\br
  T_0\, e^{\mu_{\pm}\,V_{\pm}(\g)}\,T_0^{-1} &=& \exp\left[e^{\mp\G}\mu_{\pm}\,V_{\pm}(\g)\right] \cong 1 + \mu_{\pm}e^{\mp\G}\,V_{\pm}(\g),
\er 
with $\G =  i m \left(\xi \g + \g^{-1}\eta \right) $. This construction corresponds to the Fubini-Veneziano vertex operator
which satisfies
\br
 V_+(\g_1)\,V_+(\g_2) \to 0, \qquad V_-(\g_1)\,V_-(\g_2) \to 0, \qquad \rm{as} \quad \g_1 \to \g_2 .
\er 
In general, the N-soliton solution is obtained taking $\rho =e^{\mu_1\,V(\g_1)}e^{\mu_2\,V(\g_2)}\cdots e^{\mu_N\,V(\g_N)}$, being $\mu_k$ some arbitrary parameters and the vertex functions satisfy the following commutation relation
\br
 \big[E^{(n)}, V(\g_k) \big] &=& f(n,\g_k)\,V(\g_k)\,.
\er

\section{B\"acklund Solutions}

In this section, we want to discuss the type of solutions derived from the auto-B\"acklund transformations (\ref{s3e47})-(\ref{s3e54}). Particularly, we are interested to show that these solutions are in totally consistency with  the ones given by the dressing method. Firstly, we noted that there is a closed relation between the field $\chi_+$ appearing in the dressing procedure and the defect contribution to the number of particle conserved quantity $N_D$. In fact, from eqs. (\ref{eq2.47}) it follows that
\br
{{\pa \chi_+}\o {\pa t}} ={{ig}\o {4}} (\phi_1^{\dagger} \phi_1 - \phi_2^{\dagger} \phi_2)
\label{ref1}
\er
Comparing with  eqn. (\ref{3.13}) we can take,
\br
 N_D &=& \frac{4i}{g}\(\chi_+^{(1)} - \chi_+^{(2)}\)\bigg|_{x=0},
 \label{ref2}
\er
where $\chi_+^{(i)}$ is given in general by eqn. (\ref{s2e2.58}). So, this relation gives us a connection between the dressing solutions and the B\"acklund solutions.

\subsection{Vacuum\,-\,one-soliton solution}

In order to investigate this connection, we will consider the situation of performing the B\"acklund transformation starting from the vacuum solution, 
\br
 \p_1^{(1)} = \p_2^{(1)} = \p_1^{\da(1)} = \p_2^{\da(1)} = 0, \qquad \chi_+^{(1)} =\frac{i\pi}{2}.
\er
From the explicit form of $\chi_+^{(2)}$ for one-soliton solution given by (\ref{s2e2.69}) and using the B\"acklund transformations (\ref{s3e47})-(\ref{s3e54}), we obtained 
\br
 X &=& \sqrt{\frac{m}{g}} \left[{- i {\mu_1^{(2)}a^{-1} \,e^{-\G_2}}   \over \[1+|\O_2|^2 e^{-4\G_2}\]^{\frac{1}{2}}    }\right], \qquad X^\da\,=\,\sqrt{\frac{m}{g}} \left[ {{i\mu_1^{*(2)}a^{-1} \,e^{-\G_2}}   \over \[1+|\O_2|^2 e^{-4\G_2}\]^{\frac{1}{2}} }\right],
\er
and 
\br
 \phi_1^{(2)} &=& \sqrt{\frac{m}{g}} \left[\frac{i\mu_1^{(2)} a^{-1} e^{-\G_2}}{1+\O_2 e^{-2\G_2}}\right],\qquad\quad\,\,\,
  \phi_2^{(2)} = \sqrt{\frac{m}{g}} \left[\frac{\mu_1^{(2)}  e^{-\G_2}}{1-\O_2 e^{-2\G_2 }}\right],\label{s3e65}\\[0.3cm]
   \phi_1^{\dagger(2)} &=& -\sqrt{\frac{m}{g}} \left[\frac{i\mu_1^{*(2)} a^{-1}  e^{-\G_2}}{1-\O_2 e^{-2\G_2 }}\right],\qquad\quad\,\,\,
  \phi_2^{\dagger(2)} = \sqrt{\frac{m}{g}} \left[\frac{\mu_1^{*(2)}e^{-\G_2}}{1+\O_2 e^{-2\G_2 }}\right].
\er
where $\mu_1^{(2)}$ is an arbitrary constant. Thus, we have found exactly the one-soliton solution for the BTM firstly obtained by the Dressing method, with $\g_2 = ia^{-1}$. Then, it shows that our B\"acklund transformation are compatible not only with the integrability of the Bosonic Thirring model in the presence of a jump-defect, but with the soliton solutions obtained by dressing method, too.

\subsection{One-soliton \,-\, one-soliton Solution}

We are now interested in investigating the situation of one soliton approaches a defect and on the other side of it we have an outgoing soliton. Then, we consider the form of the one-soliton solution given by (\ref{s2e70}),
\br
  \phi_1^{(k)} &=& \sqrt{\frac{m}{g}} \left[\frac{\mu_1^{(k)} \g_k e^{-\G_k}}{1+\O_k e^{-2\G_k}}\right], \qquad \phi_2^{(k)} = \sqrt{\frac{m}{g}} \left[\frac{\mu_1^{(k)}  e^{-\G_k}}{1-\O_k e^{-2\G_k}}\right],\qquad k=1,2,
\er
where $\mu_1^{(k)}$ is a parameter corresponding to each side around the defect. From (\ref{s3e47}) we find
\br
 \mu_1^{(1)}\g_1  e^{-\G_1} + \mu_1^{(2)}\g_2  e^{-\G_2} -\mu_1^{(1)} \g_1 \O_2 \,e^{-\G_1-2\G_2} -\mu_1^{(2)} \g_2 \O_1 \,e^{-\G_2-2\G_1} = \nonumber \\
- ia^{-1} \mu_1^{(1)} e^{-\G_1} + ia^{-1} \mu_1^{(2)} e^{-\G_2} - ia^{-1} \mu_1^{(1)}\O_2 \,e^{-\G_1-2\G_2} + ia^{-1}\mu_1^{(2)} \O_1 \,e^{-\G_2 -2\G_1},
\er
with $a$ being the B\"acklund parameter. This relation implies the following conditions over the parameters,
\br
 \g_1 = \g_2 = \g, \qquad  \mu_1^{(1)} \,=\, \[\frac{\s - \g}{\s + \g }\] \mu_1^{(2)},\qquad \O_1 = \[\frac{\s - \g}{\s+\g}\]^2 \,\O_2 ,
\er
where we have defined $\s=ia^{-1}$. Then, an important point that can be noted is that the jump-defect preserves the soliton velocity and the only effect of the interaction soliton-defect is a phase shift. In addition, we also have that the limiting cases when $\mu_1^{(1)}=0$ or $\mu_1^{(2)}=0$ do exist, and correspond to the situation where $\g = |\s|$ with $a>0 $, and $\g=|\s|$ with $a<0$ respectively. Clearly, these cases indicate creation and absorption of the soliton. As $a\to\infty$, the parameter $\mu_1^{(1)}\to -\mu_1^{(2)}$, which means that if the defect parameter is large the soliton will invert its shape. As $a\to 0$, we obtain $\mu_1^{(1)} = \mu_1^{(2)}$, indicating that there is not defect and the soliton shape is preserved as expected.  Some of these features have already been found for several integrable models with defects \cite{Bow2, Lean1}.

\subsection{One-soliton\,-\,two-soliton Solution}
In this case we want to examine the situation when one-soliton is incoming and two-solitons are outgoing. From Eqs.(\ref{A7}) and (\ref{s2e2.73}), we can write the one-soliton solution with para\-me\-ter $\g_1$ in terms of the tau-functions as follows,
\br
 \t_0^{(1)} &=& 1 - \frac{\mu_1^{(1)}\mu_2^ {(1)}}{4}\,e^{-2\G_1}, \qquad \t_1^{(1)}\,=\,1 + \frac{\mu_1^{(1)}\mu_2^ {(1)}}{4}\,e^{-2\G_1},\qquad  \t_2^{(1)} \,=\, \mu_1^{(1)}\g_1\,e^{-\G_1}, \quad\mbox{} \label{eq3.68}\\
\t_3^{(1)} &=& \mu_2^{(1)}\,e^{-\G_1},\qquad \t_4\,=\, \mu_1^{(1)}\,e^{-\G_1}, \qquad \t_5^{(1)}\,=\,\frac{\mu_2^{(1)}}{\g_1}\,e^{-\G_1}.\label{eq3.69}
\er
In addition, we consider the two-soliton solution (\ref{A.33})-(\ref{A.38}), with parameters $\g_1 = -\g_2$ and $\g_3 = -\g_4$. From the B\"acklund relation (\ref{eq3.14}) we obtain the following equation in terms of 
the tau-functions,
\br
 \[\t_2^{(1)}\t_0^{(2)} + \t_2^{(2)}\t_0^{(1)} \] - \s \[\t_4^{(2)}\t_1^{(1)} - \t_4^{(1)}\t_1^{(2)} \]\,=\,0, \label{eq3.71}
\er
and from Eq. (\ref{s3e47}) we get,
\br
 \[\t_3^{(1)}\t_1^{(2)} + \t_3^{(2)}\t_1^{(1)} \] + \s \[\t_5^{(2)}\t_0^{(1)} - \t_5^{(1)}\t_0^{(2)} \]\,=\,0, \label{eq4.13}
\er
where $\s=ia^{-1}$ is the parameter associated to the B\"acklund transformation. It is worth mentioning that the Eqs. (\ref{eq3.71}) and (\ref{eq4.13}) hold for all B\"acklund transitions and not only for the present case, which are a direct consequence of the expressions (\ref{s2e2.58})-(\ref{A4}). After some calculations, we found that the equation (\ref{eq3.71}) is satisfied provided the following relations hold,
\br
  \s=\g_3, \qquad \mu_1^{(2)} = \[\frac{\g_3+\g_1}{\g_3-\g_1}\] \mu_1^{(1)}, \qquad \mu_2^{(2)} = \[\frac{\g_3+\g_1}{\g_3-\g_1}\] \mu_2^{(1)}.
\er
Hence, the parameter of the B\"acklund transformation must to be exactly the second para\-me\-ter $\g_3$ of the two-soliton solution. Moreover, note that there are no conditions imposed over the parameters $\mu_3^{(2)}$ and $\mu_4^{(2)}$ because of our choice of the parameter $\g_1$ as the parameter for the one-soliton solution. In fact, if we choose $\g_3$ as the parameter for the one-soliton solution in (\ref{eq3.68})-(\ref{eq3.69}) we get the following conditions from the Eq.(\ref{eq3.71}),
\br
\s = \g_1, \qquad \mu_3^{(2)} =\[\frac{\g_3+\g_1}{\g_3-\g_1}\] \mu_1^{(1)}, \qquad \mu_4^{(2)}\,=\,\[\frac{\g_3+\g_1}{\g_3-\g_1}\]\mu_2^{(1)}.
\er
In this case, the parameter of the B\"acklund transformation must to be $\g_1$ and $\mu_1^{(2)}$ and $\mu_2^{(2)}$ are arbitraries. These results can be put together on a compact form by taking $\g$ as the one-soliton solution parameter and $\s$ the B\"acklund parameter, 
\br
 \mu_1^{(2)}=-\mu_3^{(2)} =\[\frac{\s+\g}{\s-\g}\]\mu_1^{(1)}, \qquad \mu_2^{(2)} = -\mu_4^{(2)} =\[\frac{\s+\g}{\s-\g}\]\mu_2^{(1)},
\er
where $\g$ and $\s$ can take the values $\{\g_1,\g_3\}$. In others words, these results is providing us of an indirectly evidence of the \emph{permutability theorem} of the B\"acklund transformations.

\subsection{Two soliton-two soliton Solution}

For this time we consider the two soliton solution given explicitely by the Eqs. (\ref{A.33})-(\ref{A.38}) for both sides of the defect. Then, the two B\"acklund relations (\ref{eq3.71}) and (\ref{eq4.13}) are satisfied provided the following realtions between the parameters hold,
\br
 \mu_1^{(1)} =\[\frac{\s-\g_1}{\s+\g_1}\]\mu_1^{(2)}, \qquad  \mu_3^{(1)} =\[\frac{\s-\g_3}{\s+\g_3}\]\mu_3^{(2)},\\
 \mu_2^{(1)} =\[\frac{\s-\g_1}{\s+\g_1}\]\mu_2^{(2)}, \qquad  \mu_4^{(1)} =\[\frac{\s-\g_3}{\s+\g_3}\]\mu_4^{(2)},
\er
defining the corresponding phase shifts.


\section{Defect Matrix for the Bosonic Thirring Model}

In order to calculate the defect matrix, it is necessary to introduce the modified Lax pair defined in two overlapping regions containing the defect. Following \cite{Corr1,Bow1}, the space can be splitted by introducing two points $0<a<b$ which define two regions : on the left $R^{(1)}$, $-\infty < x < b$, and on the right $R^{(2)}$ , $a<x<\infty$. Then, the modified Lax pair can be written in a general form, as follows
\br
 {\hat A}_t^{(1)} &=& A_t^{(1)} + \th(x-a) \[ F_1\(\p_1^{(1)},\p_1^{\da(2)},X\)\s_+^{(0)} -     F_1^{\da}\(\p_1^{\da(1)},\p_1^{\da(2)},X^\da\)\s_-^{(+1)}\],\qquad \mbox{}\\
  {\hat A}_x^{(1)} &=& A_x^{(1)}\,\th(a-x),\\[0.3cm]
 {\hat A}_t^{(2)} &=& A_t^{(2)} + \th(b-x) \[ F_2\(\p_2^{(1)},\p_2^{(2)},X\) \s_+^{(-1)} -     F_2^{\da}\(\p_2^{\da(1)},\p_2^{\da(2)},X^\da\)\s_-^{(0)}\],\qquad \mbox{}\\
  {\hat A}_x^{(2)} &=& A_x^{(2)}\,\th(x-b).
\er
where $A_t$ and $A_x$ are related with the connections introduced in (\ref{lax}) by the following simple relations,
\br
  A_t^{(p)} &=& \frac{1}{2}\(A_{\xi}^{(p)}+A_{\eta}^{(p)}\), \quad A_x^{(p)} \,=\, \frac{1}{2}\(A_{\xi}^{(p)}-A_{\eta}^{(p)}\),\quad p=1,2,
\er
and the functions $F_i$ and $F^{\da}_i$ are given explicitly by
\br
 F_1\(\p_1^{(1)},\p_1^{\da(2)},X\) &=& X -  \p_1^{(1)} \exp\[\frac{ig N_D}{4}\] - \p_1^{(2)} \exp\[\frac{-ig N_D}{4}\]  ,\\
 F_1^{\da}\(\p_1^{\da(1)},\p_1^{\da(2)},X^\da\) &=& X^\da -  \p_1^{\da(1)} \exp\[\frac{-ig N_D}{4}\] - \p_1^{\da(2)} \exp\[\frac{ig N_D}{4}\],\\
 F_2\(\p_2^{(1)},\p_2^{(2)},X\) &=& X- ia^{-1} \[ \p_2^{(2)} \exp\[\frac{ig N_D}{4}\] - \p_2^{(1)} \exp\[\frac{-ig N_D}{4}\]\],\\
  F_2^\da\(\p_2^{\da(1)},\p_2^{\da(2)},X^\da\) &=& X^\da + ia^{-1}\[\p_2^{\da(2)} \exp\[\frac{-ig N_D}{4}\] - \p_2^{\da(1)} \exp\[\frac{ig N_D}{4}\] \].
\er
These modified Lax pair allow us to derive the equations of motion for each region  after applying the zero curvature condition, and the functions $F_i$ and $F_i^{\da}$ provide us the defect relations or auto-B\"acklund transformations in the same way. So, within the overlap region the modified Lax pair are related by the gauge transformation,
\br
 i\pa_t K &=& K{\hat A_t}^{(2)} - {\hat A}_t^{(1)}K.\label{6.1}
\er
\noindent 
Taking into account the explicit symmetric form of the Lax pair of the Bosonic Thirring model, we propose  the following ansatz for the form of the defect matrix $K$ in the $\l$-expansion, 
\br
 K &=& K_{-1} +  K_0 + K_1 ,
\er
where $K_i$ corresponds to an element of grade $\l^{i}$. Implementing the same procedure used for the Grassmaniann case, we compute the explicit form for the defect matrix $K$, and after some algebra we found the following solution: 
\br
 K &=& \left[
   \begin{array}{cc} -\sqrt{\frac{m}{g}}\[ \l e^{-\frac{ig N_D}{4}} - \l^{-1} (ia^{-1})e^{\frac{ig N_D}{4}} \]  &\l^{0} X \\[0.3cm]
   -\l^{0} X^{\da} & \sqrt{\frac{m}{g}}\[\l e^{\frac{ig N_D}{4}} + \l^{-1} (ia^{-1})e^{-\frac{ig N_D}{4}} \]
   \end{array}
 \right],\qquad \mbox{}
\er
where we have used the B\"acklund transformation (\ref{s3e47})-(\ref{s3e54}) and the equations of motion (\ref{eq2.29})-(\ref{eq2.32}).

\section{Conclusions}

In conclusion, by expressing the B\"acklund transformation for the Grassmanian Thirring model in terms of the defect number of particles we have been able to generalize this transformation to the Bosonic Thirring model. By obtaining the solutions through a dressing method which  determines directly the defect number of particles we have been able to verify the B\"acklund transformation for several simple transitions, i.e. vacuum-one-soliton, one-soliton-one-soliton, one-soliton-two-solitons, and two-solitons-two-solitons.


\vskip 1cm
 \noindent
{\bf Acknowledgements} \\
\vskip .1cm \noindent
{ARA and LHY thank Fapesp,   JFG and AHZ thank CNPq for partial
support.}
\bigskip

\appendix

\section{The one-soliton Solution}

Using the highest weight representation of $\hat{s}l(2)$ we obtain the
one-soliton solution from the vacuum configuration, as follows\footnote{For this solution $c$
is required to be $c=1$.} 
\br
  e^{(\nu_- -\nu_+)} &=& \t_0 ,\qquad  e^{-2\chi_+}\,=\, - { \t_1\over\t_0},\label{s2e2.58}\\
  \phi_1 &=&\sqrt{\frac{m}{g}}\,{\t_2\over \t_1}, \qquad \phi_2 \,=\, \sqrt{\frac{m}{g}}\,{\t_4 \over \t_0},\\
  \phi_1^{\dagger} &=& \sqrt{\frac{m}{g}}\,{\t_3 \over \t_0}, \,\,\,\qquad \phi_2^{\dagger}\,=\,\sqrt{\frac{m}{g}}\,{\t_5\over\t_1},
\er
where we have introduced the tau-functions
\br
\t_0 &=& \langle \l_0 |G |\l_0 \rangle, \qquad \t_2 = \langle \l_0 | \s_-^{(+1)}G   |\l_0 \rangle , \qquad \t_4\,=\,\langle \l_1 | G\s_-^{(0)}   |\l_1 \rangle, \nonu \\
\t_1 &= & \langle \l_1 |  G  |\l_1 \rangle, \qquad  \t_3 = \langle \l_1 | \s_+^{(0)}G   |\l_1 \rangle, \,\,\,\qquad  
\t_5\,=\,\langle \l_0 | G  \s_+^{(-1)} |\l_0 \rangle, \label{A4}
\er
and where $G= T_0 \,\rho\, T_0^{-1}$ and $\rho = e^{V}$. Firstly, we can noticed that there are two possible 
solutions corresponding to the choice of $V = \mu_1 V_+(\g_1)$,  given by 
\br
 \nu_+ = \nu_- ,\quad \chi_+ = \frac{i \pi}{2}, \quad \phi_1 = \sqrt{\frac{m}{g}} \mu_1 \g e^{-\G_1}, \quad \phi_2 = \sqrt{\frac{m}{g}} \mu_1 e^{-\G_1}, \quad \phi_1^{\dagger} = \phi_2^{\dagger}=0,
\er 
and by choosing $V= \mu_2 V_-(\g_2)$, we obtain
\br
 \nu_+ = \nu_- ,\quad \chi_+ = \frac{i \pi}{2}, \quad \phi_1 = \phi_2=0,\quad \phi_1^{\dagger} = \sqrt{\frac{m}{g}} \mu_2 e^{\G_2}, \quad \phi_2^{\dagger} = \sqrt{\frac{m}{g}} \frac{\mu_2}{\g} e^{\G_2}.
\er 
In our case, these solutions are not interesting because of the
inconsistency with the interpretation of the dagger fields
$\phi^{\dagger}$'s as the corresponding complex conjugate of the
fields $\phi$'s. Then, we construct the one-soliton solution of the
system using the fact that $V_+(\g)$ and $V_-(-\g)$ have the same
eigenvalue. In fact,  by choosing $ \rho = e^{\mu_1 V_+(\g_1)}
e^{\mu_2 V_-(\g_2)}$ and computing the matrix elements we get the
following solution, 
\br
 \t_0 = 1 + \mu_1 \mu_2 \left[ \frac{\g_1\g_2}{(\g_1 -\g_2)^2} \right]e^{-\G_1 + \G_2},\qquad 
 \t_1= 1+\mu_1\mu_2 \left[ \frac{\g_1}{\g_1 - \g_2} \right]^2 e^{-\G_1 + \G_2},\nonu  \\
 \t_0^{(-)}= \mu_1 \g_1 e^{-\G_1}, \qquad \t_1^{(-)}= \mu_1  e^{-\G_1}, \qquad 
 \t_1^{(+)} = \mu_2  e^{\G_2}, \qquad  \t_0^{(+)} = {{\mu_2}\over {\g_2}} \,e^{\G_2}\label{A7}
 \er
where $\G_k = im\left(\xi \g_k + \g_k^{-1} \eta \right)$. Then, we
are interested in the case where $\phi_k^{\dagger}$ corresponds to
the complex conjugate of $\phi_k$, i.e., in the limit $\g_2 \to
-\g_1$,  which provides a suitable one-soliton solution for the
Bosonic Thirring model. The result is 
\br
 e^{(\nu_- -\nu_+)} &=& 1-  \O\,e^{-2\G_1}, \qquad \qquad \quad \,\,
 e^{-2\chi_+} = -\left[\frac{1+ \O e^{-2\G_1}}{1 -  \O\,e^{-2\G_1}}\right] , \label{s2e2.69}\\[0.2cm]
 \phi_1 &=& \sqrt{\frac{m}{g}} \left[\frac{\mu_1 \g_1 e^{-\G_1}}{1+\O e^{-2\G_1}}\right],\qquad
  \phi_2 = \sqrt{\frac{m}{g}} \left[\frac{\mu_1  e^{-\G_1}}{1-\O e^{-2\G_1 }}\right],\label{s2e70}\\
   \phi_1^{\dagger} &=& \sqrt{\frac{m}{g}} \left[\frac{\mu_2  e^{-\G_1}}{1-\O e^{-2\G_1 }}\right],\qquad
  \phi_2^{\dagger} = -\sqrt{\frac{m}{g}} \frac{\mu_2}{\g_1}\left[\frac{e^{-\G_1}}{1+\O e^{-2\G_1 }}\right],\label{s2e2.73}
\er 
where we have introduced the parameter $\O = \frac{\mu_1
\mu_2}{4}$.  Considering $m$ and $g$ to be real, and $\g_1$ purely
imaginary, from (\ref{s2e2.69}-\ref{s2e2.73}) one gets that the parameters
$\mu_+$ and $\mu_-$ must satisfy the following relation, 
\br
 \mu_2 &=& -\g_1 \mu_1^{*},
\er
We can also notice that for an appropriated choice of the parameters, it is possible to show the equivalence with the one-soliton solution found by Orfanidis\cite{Orf}.

\section{The two-soliton Solution}
Now let us show that the two-soliton solution can be also calculated
from the vacuum solution (\ref{eq2.38}) using the dressing
transformation. We will do it using only the algebraic properties of
the affine Lie algebra $\hat{s}l(2)$. According to the approach
above, there is an element $\rho$ in the group satisfying
(\ref{eq2.50}). Consider the constant group element as 
\br
 \rho = e^{\mu_1 V_+(\g_1)} e^{\mu_2 V_-(\g_2)} e^{\mu_3 V_+(\g_3)} e^{\mu_4 V_-(\g_4)}.
\er 

\noindent The explicit form for the solution is calculated by
computing the following matrix elements, 
\br
\langle \l_0 | V_+(\g_1) V_-(\g_2) |\l_0\rangle &=&\langle \l_0 |  V_-(\g_2)V_+(\g_1) |\l_0\rangle \,=\, \frac{\g_1 \g_2}{(\g_1 -\g_2)^2}, \\
\langle \l_1 | V_+(\g_1) V_-(\g_2) |\l_1\rangle &=&\langle \l_1 |
V_-(\g_2)V_+(\g_1)  |\l_1\rangle \,=\,
\frac{\g_1^2}{(\g_1-\g_2)^2}. \er

\noindent In addition, one has 
\br
 \langle \l_0 | \s_-^{(+1)} V_+(\g_1) |\l_0\rangle &=& \g_1,\qquad  \langle \l_1 | \s_+^{(0)} V_-(\g_1) |\l_1\rangle\,=\, 1,\\
 \langle \l_0 | V_-(\g_2)\s_+^{(-1)} |\l_0\rangle &=&\frac{1}{\g_2}, \qquad  \langle \l_1 | V_+(\g_1)\s_-^{(0)} |\l_0\rangle \,=\,1,
\er 
and 
\br
\hspace{-0.5cm} \langle \l_0 | V_+(\g_1)V_-(\g_2)V_+(\g_3)V_-(\g_4) |\l_0\rangle &=& \left[\frac{\g_1\g_2\g_3\g_4 (\g_1-\g_3)^2(\g_2-\g_4)^2}{(\g_1-\g_2)^2(\g_3-\g_4)^2(\g_1-\g_4)^2(\g_2-\g_3)^2}\right], \quad \,\mbox{}\\[0.2cm]
\hspace{-0.5cm} \langle \l_1 | V_+(\g_1)V_-(\g_2)V_+(\g_3)V_-(\g_4) |\l_1\rangle &=& \left[\frac{\g_1^2\g_3^2(\g_1-\g_3)^2(\g_2-\g_4)^2}{(\g_1-\g_2)^2(\g_3-\g_4)^2(\g_2-\g_3)^2(\g_1-\g_4)^2}\right], \quad \,\mbox{}\\[0.2cm]
\hspace{-0.5cm} \langle \l_0 |\s_-^{(+1)} V_+(\g_1)V_-(\g_2)V_+(\g_3) |\l_0\rangle &=& \left[\frac{\g_1\g_2\g_3(\g_1-\g_3)^2}{(\g_1-\g_2)^2(\g_2-\g_3)^2}\right], \quad \,\mbox{}\\[0.2cm]
\hspace{-0.5cm} \langle \l_0 |\s_-^{(+1)} V_+(\g_1)V_+(\g_3)V_-(\g_4) |\l_0\rangle &=& \left[\frac{\g_1\g_3\g_4(\g_1-\g_3)^2}{(\g_3-\g_4)^2(\g_1-\g_4)^2}\right], \quad \,\mbox{}\\[0.2cm]
\hspace{-0.5cm} \langle \l_1 |\s_+^{(0)} V_-(\g_2)V_+(\g_3)V_-(\g_4) |\l_1\rangle &=& \left[\frac{\g_3^2(\g_2-\g_4)^2}{(\g_2-\g_3)^2(\g_3-\g_4)^2} \right], \quad \,\mbox{}\\[0.2cm]
\hspace{-0.5cm} \langle \l_1 |\s_+^{(0)} V_+(\g_1)V_-(\g_2)V_-(\g_4) |\l_0\rangle &=& \left[\frac{\g_1^2(\g_2-\g_4)^2}{(\g_1-\g_2)^2(\g_1-\g_4)^2} \right], \quad \,\mbox{}\\[0.2cm]
\hspace{-0.5cm} \langle \l_1 | V_+(\g_1)V_+(\g_3)V_-(\g_4) \s_-^{(0)}|\l_1\rangle &=& \frac{\g_4^2(\g_1 -\g_3)^2}{(\g_1 -\g_4)^2(\g_3-\g_4)^2}, \quad \,\mbox{}\\[0.2cm]
\hspace{-0.5cm} \langle \l_1 | V_+(\g_1)V_-(\g_2)V_+(\g_3)\s_-^{(0)} |\l_1\rangle &=& \frac{\g_2^2(\g_1 -\g_3)^2}{(\g_1 -\g_2)^2(\g_2-\g_3)^2}, \quad \,\mbox{}\\[0.2cm]
\hspace{-0.5cm} \langle \l_0| V_-(\g_2)V_+(\g_3)V_-(\g_4)\s_+^{(-1)} |\l_0\rangle &=& \frac{\g_3^3(\g_2 -\g_4)^2}{\g_2\g_4(\g_2 -\g_3)^2(\g_3-\g_4)^2}, \quad \,\mbox{}\\[0.2cm]
\hspace{-0.5cm} \langle \l_0| V_+(\g_1)V_-(\g_2)V_-(\g_4)\s_+^{(-1)}
|\l_0\rangle &=&  \frac{\g_1^3(\g_2 -\g_4)^2}{\g_2\g_4(\g_1
-\g_2)^2(\g_1-\g_4)^2}, \quad \,\mbox{} 
\er

\noindent So, we obtain the following results, 
\br
 \t_0 &=&  1 + \mu_1\mu_2 e^{-\G_1+\G_2}\left[\frac{\g_1\g_2}{(\g_1-\g_2)^2}\right] + \mu_1\mu_4 e^{-\G_1+\G_4}\left[\frac{\g_1\g_4}{(\g_1-\g_4)^2}\right]\nonumber \\
 &+&   \mu_2\mu_3 e^{\G_2-\G_3}\left[\frac{\g_2\g_3}{(\g_2-\g_3)^2}\right] + \mu_3\mu_4 e^{-\G_3+\G_4}\left[\frac{\g_3\g_4}{(\g_3-\g_4)^2}\right] \nonumber\\
 &+&    \mu_1\mu_2\mu_3\mu_4 e^{-\G_1+\G_2-\G_3+\G_4}\left[\frac{\g_1\g_2\g_3\g_4 (\g_1-\g_3)^2(\g_2-\g_4)^2}{(\g_1-\g_2)^2(\g_3-\g_4)^2(\g_1-\g_4)^2(\g_2-\g_3)^2}\right],
\er

\br
 \t_1 &=&  1 + \mu_1\mu_2 e^{-\G_1+\G_2}\left[\frac{\g_1^2}{(\g_1-\g_2)^2}\right] + \mu_1\mu_4 e^{-\G_1+\G_4}\left[\frac{\g_1^2}{(\g_1-\g_4)^2}\right]\nonumber \\
 &+&   \mu_2\mu_3 e^{\G_2-\G_3}\left[\frac{\g_3^2}{(\g_2-\g_3)^2}\right] + \mu_3\mu_4 e^{-\G_3+\G_4}\left[\frac{\g_3^2}{(\g_3-\g_4)^2}\right] \nonumber\\
 &+&    \mu_1\mu_2\mu_3\mu_4 e^{-\G_1+\G_2-\G_3+\G_4} \left[\frac{\g_1^2\g_3^2(\g_1-\g_3)^2(\g_2-\g_4)^2}{(\g_1-\g_2)^2(\g_3-\g_4)^2(\g_2-\g_3)^2(\g_1-\g_4)^2}\right], \\[0.4cm]
 \t_2 &=& \mu_1 \g_1 e^{-\G_1} + \mu_3 \g_3 e^{-\G_3} +\mu_1\mu_2\mu_3 e^{-\G_1+\G_2-\G_3}\left[\frac{\g_1\g_2\g_3(\g_1-\g_3)^2}{(\g_1-\g_2)^2(\g_2-\g_3)^2}\right]\nonumber \\
 &+& \mu_1\mu_3\mu_4 e^{-\G_1-\G_3+\G_4}\left[\frac{\g_1\g_3\g_4(\g_1-\g_3)^2}{(\g_3-\g_4)^2(\g_1-\g_4)^2}\right],\\[0.4cm]
 \t_3 &=& \mu_2 e^{\G_2} + \mu_4 e^{\G_4} + \mu_2\mu_3\mu_4 e^{\G_2-\G_3+\G_4} \left[\frac{\g_3^2(\g_2-\g_4)^2}{(\g_2-\g_3)^2(\g_3-\g_4)^2} \right]\nonumber\\
 &+& \mu_1\mu_2\mu_4 e^{-\G_1+\G_2+\G_4}\left[\frac{\g_1^2(\g_2-\g_4)^2}{(\g_1-\g_2)^2(\g_1-\g_4)^2} \right],\\[0.4cm]
  \t_4 &=& \mu_1 e^{-\G_1} +\mu_3 e^{-\G_3} +\mu_1\mu_2\mu_3 e^{-\G_1+\G_2-\G_3} \left[     \frac{\g_2^2(\g_1 -\g_3)^2}{(\g_1 -\g_2)^2(\g_2-\g_3)^2}\right] \nonumber \\
 &+& \mu_1\mu_3\mu_4 e^{-\G_1-\G_3+\G_4}\left[ \frac{\g_4^2(\g_1 -\g_3)^2}{(\g_1 -\g_4)^2(\g_3-\g_4)^2}\right],\\[0.4cm]
\t_5 &=& \frac{\mu_2}{\g_2} e^{\G_2}+ \frac{\mu_4}{\g_4} e^{\G_4} +\mu_2\mu_3\mu_4 e^{\G_2 -\G_3 + \G_4} \left[\frac{\g_3^3(\g_2 -\g_4)^2}{\g_2\g_4(\g_2 -\g_3)^2(\g_3-\g_4)^2}\right]\nonumber\\
&+&\mu_1\mu_2\mu_4 e^{-\G_1 +\G_2 + \G_4} \left[\frac{\g_1^3(\g_2
-\g_4)^2}{\g_2\g_4(\g_1 -\g_2)^2(\g_1-\g_4)^2} \right]. 
\er\\
We can check that these tau-functions satisfy the equations (\ref{eq2.29})-(\ref{eq2.32}) and (\ref{eq2.47}) for any values of the parameters $\mu_k$ and $\g_k$, with $k=1,...,4$. Now, taking the limit $\g_2\to -\g_1$ and $\g_4\to -\g_3$, we get the two-soliton solution for the Bosonic Thirring model. The tau-functions  become, 
\br
 \t_0 &=& 1 - \frac{\mu_1\mu_2}{4} e^{-2\G_1} -  \frac{\mu_3\mu_4}{4} e^{-2\G_3}
 -  (\mu_1\mu_4+\mu_2\mu_3) e^{-(\G_1+\G_3)}\left[\frac{\g_1\g_3}{(\g_1+\g_3)^2}\right] \nonumber \\
 &+& \frac{1}{16}(\mu_1\mu_2\mu_3\mu_4) e^{-2(\G_1+\G_3)}\left[\frac{\g_1-\g_3}{\g_1+\g_3}\right]^4 \label{A.33}\\[0.4cm]
 \t_1 &=& 1 + \frac{\mu_1\mu_2}{4} e^{-2\G_1} +  \frac{\mu_3\mu_4}{4} e^{-2\G_3} + (\g_1^2 \mu_1\mu_4 +\g_3^2 \mu_2\mu_3) e^{-(\G_1+\G_3)}\left[\frac{1}{(\g_1+\g_3)^2}\right]\\
 &+&\frac{1}{16}(\mu_1\mu_2\mu_3\mu_4) e^{-2(\G_1+\G_3)}\left[\frac{\g_1-\g_3}{\g_1+\g_3}\right]^4 \nonumber\\[0.4cm]
 \t_2 &=& \mu_1 \g_1 e^{-\G_1} + \mu_3 \g_3 e^{-\G_3} -\frac{1}{4}\mu_1\mu_2\mu_3 e^{-(2\G_1+\G_3)}\left[\frac{\g_3(\g_1-\g_3)^2}{(\g_1+\g_3)^2}\right]\nonumber\\
 &-&\frac{1}{4}\mu_1\mu_3\mu_4 e^{-(\G_1+2\G_3)}\left[\frac{\g_1(\g_1-\g_3)^2}{(\g_1+\g_3)^2}\right],\\[0.4cm]
\t_3 &=&\mu_2 e^{-\G_1} + \mu_4 e^{-\G_3} +\frac{1}{4} \mu_2\mu_3\mu_4 e^{-(\G_1+2\G_3)} \left[\frac{\g_1-\g_3}{\g_1+\g_3} \right]^2\nonumber\\
 &+& \frac{1}{4}\mu_1\mu_2\mu_4 e^{-(2\G_1+\G_3)}\left[\frac{\g_1-\g_3}{\g_1+\g_3} \right]^2,\\[0.4cm]
 \t_4 &=&\mu_1 e^{-\G_1} +\mu_3 e^{-\G_3} +\frac{1}{4}\mu_1\mu_2\mu_3 e^{-(2\G_1+\G_3)} \left[\frac{\g_1 -\g_3}{\g_1+\g_3}\right]^2 \nonumber \\
 &+& \frac{1}{4}\mu_1\mu_3\mu_4 e^{-(\G_1+2\G_3)}\left[ \frac{\g_1 -\g_3}{\g_1 +\g_3}\right]^2,\\[0.4cm]
 \t_5 &=&-\frac{\mu_2}{\g_1} e^{-\G_1} - \frac{\mu_4}{\g_3} e^{-\G_3} +\frac{1}{4}\mu_2\mu_3\mu_4 e^{-(\G_1+2\G_3)} \left[\frac{(\g_1 -\g_3)^2}{\g_1(\g_1+\g_3)^2}\right]\nonumber\\
&+& \frac{1}{4}\mu_1\mu_2\mu_4 e^{-(2\G_1 +\G_3)} \left[\frac{(\g_1
-\g_3)^2}{\g_3(\g_1 +\g_3)^2} \right].\label{A.38} 
\er

\noindent 
Considering that $m$ and $g$ must be real constants, there are two possibilities in order to $\p_k^\da$ does corresponds to the complex conjugate of $\p_k$. First one corresponds to the case of $\g_1$ and
$\g_3$ to be purely imaginary numbers, and the parameters $\mu_k$ satisfyng the following conditions,
\br
 \mu_2 & =& -\g_1 \mu_1^*, \qquad \quad \mu_4= -\g_3\mu_3^*.
\er
The second possibility corresponds to the situation when $\g_3^* = -\g_1$, and as consequence $\G_3^*=\G_1$. In this case, we need that the parameters $\mu_k$ satisfy the following conditions,
\br
 \mu_4^* &=& \g_1\mu_1, \qquad \quad \mu_2^*\,=\,\g_3\mu_3. \label{eq2.104}
\er



\section{The $sl(2,1)$ affine Lie algebra notations}

Consider the $\hat{s}l(2,1)$ super Lie algebra with its generators given by
\br
 \hspace{-2cm}h_1 &=& \a_1 \cdot H \,=\,\left(\begin{array}{crc}1&0&0\\0&-1&0\\0&0&0\end{array}\right),\quad h_2\,=\,\a_2 \cdot H\,=\,\left(\begin{array}{ccc}0&0&0\\0&1&0\\0&0&1\end{array} \right),\nonumber \\
  E_{\a_1}&=&\left(\begin{array}{ccc}0&1&0\\0&0&0\\0&0&0\end{array}\right),\quad E_{-\a_1}\,=\,\left(\begin{array}{ccc}0&0&0\\1&0&0\\0&0&0\end{array}\right),\quad E_{\a_2}\,=\,\left(\begin{array}{ccc}0&0&0\\0&0&1\\0&0&0\end{array}\right) , \\
 E_{-\a_2}&=&\left(\begin{array}{ccc}0&0&0\\0&0&0\\0&1&0\end{array}\right),\quad
 E_{\a_1+\a_2}\,=\,\left(\begin{array}{ccc}0&0&1\\0&0&0\\0&0&0\end{array}\right),\quad E_{-(\a_1+\a_2)}\,=\,\left(\begin{array}{ccc}0&0&0\\0&0&0\\1&0&0\end{array}\right),\nonumber
\er\\
where $\a_1$ is a bosonic root and $\a_2$, $\a_1+\a_2$ are the fermionic roots. The affine $\hat{s}l(2,1)$ algebra is decomposed according to the \emph{grading operator}
\br
 Q &=& 2d +\frac{1}{2}h_1,
\er
where $d$ is the \emph{derivation operator} satisfying $[d,T_a^{(n)}]=nT_a^{(n)}$. Here $T_a^{(n)}$ denotes both $H_i^{(n)}$ and $E_{\a}^{(n)}$. The hierarchy is further specified by the constant grade one element $E=E^{(1)}$, as follows
\br
 E^{(2n+1)}&=& h_1^{(n+1/2)} + 2h_2^{(n+1/2)}\,=\,K_2^{(2n+1)}\,,
 \label{7.3}
\er 
here $\mu_i$ denotes the $i$-th fundamental weight. The grading operator $Q$ together with the judicious choice of $E$ decomposes the affine super Kac-Moody algebra $\hat{\lie}=\hat{s}l(2,1)$ into ${\hat\lie} = \cal{K}\oplus\cal{M}$, where the Kernel ${\cal K}=\{x\in\hat\lie|[x,E]=0\}$ of E, and its complement $\cal{M}$ are given by\\
\br
 \cal{K} &=& \{K_1^{(2n+1)}, K_2^{(2n+1)}, M_1^{(2n+1)}, M_2^{(2n)}  \},\\
 \cal{M} &=& \{F_1^{(2n+3/2)}, F_2^{(2n+1/2)}, G_1^{(2n+1/2)}, G_2^{(2n+3/2)}\},
\er
where the bosonic generators are 
\br
 M_1^{(2n+1)} &=& -\big(E_{\a_1}^{(n)}-E_{-\a_1}^{(n+1)}\big),\qquad M_2^{(2n)}\,=\,h_1^{(n)},\\
 K_1^{(2n+1)} &=& -\big(E_{\a_1}^{(n)}+E_{-\a_1}^{(n+1)}\big),\qquad K_2^{(2n+1)}\,=\,\mu_2\cdot H^{(n+1/2)}, 
\er
and the fermionic generators are
\br
 F_1^{(2n+3/2)} &=& \big(E_{\a_1+\a_2}^{(n+1/2)} - E_{\a_2}^{(n+1)}\big) + \big(E_{-(\a_1+\a_2)}^{(n+1)} - E_{-\a_2}^{(n+1/2)}\big), \\
 F_2^{(2n+1/2)} &=& -\big(E_{\a_1+\a_2}^{(n)} - E_{\a_2}^{(n+1/2)}\big) + \big(E_{-(\a_1+\a_2)}^{(n+1/2)} - E_{-\a_2}^{(n)}\big),\\
 G_1^{(2n+1/2)} &=&  \big(E_{\a_1+\a_2}^{(n)} + E_{\a_2}^{(n+1/2)}\big) + \big(E_{-(\a_1+\a_2)}^{(n+1/2)} + E_{-\a_2}^{(n)}\big),\\
 G_2^{(2n+3/2)} &=& -\big(E_{\a_1+\a_2}^{(n+1/2)} + E_{\a_2}^{(n+1)}\big) + \big(E_{-(\a_1+\a_2)}^{(n+1)} + E_{-\a_2}^{(n+1/2)}\big)
\er

\end{document}